\documentclass[12pt]{article}
 \usepackage{mathrsfs}
\usepackage{amsmath}
\usepackage{graphicx,color}
\usepackage{subfigure}
\usepackage{graphicx,wrapfig}

\global\arraycolsep=1pt \oddsidemargin .20in \evensidemargin .5in
\usepackage[Symbol]{upgreek}
\usepackage{bbm}
\usepackage{dsfont}
\usepackage{amssymb}
\usepackage{textcomp}

\newcommand{\CR}{\nonumber \\*}

\newcommand{\trace}{\hbox {Tr}~}

\DeclareMathAlphabet{\mathpzc}{OT1}{pzc}{m}{it}

\def\e{\mathfrak{e}}

\def\p{\partial}

\usepackage[colorlinks=true,backref=true,linkcolor=black,anchorcolor=black,citecolor=black,filecolor=black,menucolor=black,pagecolor=black,urlcolor=black]{hyperref}

\usepackage[Symbol]{upgreek}
\usepackage{caption}
\usepackage{bbm}
\usepackage{dsfont}
\usepackage{amssymb}
\usepackage{textcomp}

\def\e{\epsilon}

\def\z{\zeta}
\def\z{{\bar z}}

\def\be{\begin{equation}}
\def\ee{\end{equation}}
\def\bea{\begin{eqnarray}}
\def\eea{\end{eqnarray}}
\def\bdis{\begin{displaymath}}
\def\edis{\end{displaymath}}
\def\corr{$\clubsuit$}

\def\Z{\mathscr{Z}}

\begin{document}

\newcommand{\dslash}{\partial\!\!\!/}
\newcommand{\aslash}{A\!\!\!/}
\newcommand{\Dslash}{D\!\!\!\!/}
\newcommand{\pslash}{p \hspace{-1.7mm} /}
\newcommand{\kslash}{k \hspace{-1.7mm} /}
\newcommand{\bs}{b \hspace{-1.7mm} /}

\allowdisplaybreaks[1]
\renewcommand{\thefootnote}{\fnsymbol{footnote}}
\def\corr{$\spadesuit $}
\def\trefle{ $\clubsuit$}

\renewcommand{\thefootnote}{\arabic{footnote}}
\setcounter{footnote}{0}


 \def\stop{$\blacksquare$}
\begin{titlepage}
\null
\begin{flushright}
\end{flushright}
\begin{center}
{{\Large \bf
Minimizing gauge-functional for  2-d gravity and string theory
 }}
\lineskip .75em \vskip 3em \normalsize {\large Laurent
Baulieu     
\footnote {email address: baulieu@lpthe.jussieu.fr}
\\
 {\it Theoretical Division CERN}\footnote{
 CH-1211 Gen\`eve, 23, Switzerland }
 \\
 {\it LPTHE  Universit\'e Pierre et Marie Curie
 }\footnote{ 4 place Jussieu, F-75252 Paris
 Cedex 05, France.}
 }\\
\vskip 1em
 \normalsize {\large Daniel  Zwanziger}\footnote{email address:  daniel.zwanziger@nyu.edu}
\\
 {\it Physics Department,   New York University
 }\footnote{ 4 Washington Place, New York, NY 10003, USA
.}
\vskip 1 em
\end{center}
\begin{abstract}
 
 \end{abstract}
We   show the existence of a minimizing procedure for selecting a unique representative on the orbit of any given Riemann surface that contributes to the string  partition function. As it must, the procedure reduces the string path integral to a final  integration over  a particular fundamental domain,    selected by the choice of the minimizing functional. This construction  somehow   demystifies the Gribov question.

\end{titlepage}
 \def\Tr{\trace}

 \def\bz{{\bar z}}
 \def\mb{\mu^z_\bz}
 \def\mbb{\mu^\bz_z}
\def\dz{\partial _z }
\def\dzb{\partial _\bz }
\def\cz{c ^z }
\def\czb{c ^{\bz}} 
\def\Cz{C_\bz }
\def\Czb{C _{z}}
\section{Introduction}

 In this article, we describe  a  procedure for gauge-fixing the 2d-gravity gauge invariance~\cite{polyakov} with a   geometrical  meaning that is as transparent as possible.   
 The aim is to find a definition  that    escapes the Gribov question \cite{gribov} and, more precisely, to select unambiguosuly a single representative
 on the orbits of the conformal classes of metrics of Riemann surfaces.
 
 
A better  understanding of  the gauge-fixing is useful for    the predictions of string theory.
 For instance, for on-shell string  observables, the 
  singularities     at  the   boundaries  of the moduli space of     string worldsheets  are the source 
  of    infra-red divergencies of the    4-dimensional quantum field theory limit.    An unambiguous string     gauge-fixing method is certainly needed. 

 We will show the existence of a minimizing procedure for selecting a unique representative on the orbit of any given Riemann surface that contributes to the partition function, while making sure that a BRST symmetry is maintained.  The procedure selects among all representatives of the worldsheet in the Teichm\"uller space a particular fundamental domain, which is made of representatives that are at   absolute minimum distance of a reference worldsheet.  We will use the framework of  the Beltrami parametrization of   string 2d-worldsheets (and its extension for the superstring).

 The method holds for any given fixed genus. For the torus, the procedure may be tuned to select  the first fundamental domain of the Poincar\'e disk.
In this method,  the obvious inconsistencies  of the Faddeev--Popov method in the  ``conformal gauge", and possible Gribov copies are successfully eliminated.\footnote{The necessity of 
 selecting  a single representative for each orbit in a BRST invariant way,  is justified, since it provides   the safe definition of quantum observables  as the elements of the   cohomology of the BRST symmetry. }


 The Beltrami parametrization of the 2-dimensional metrics   in  string theory was   introduced  in 1986 for a  clearer   definition of     the path integral of the 2d-gravity field~\cite{baulieubelloneta}\cite{baulieubelloneta1}. It gives  a better understanding  of the factorization of left and right movers  and of the  the conformal anomaly of  string theory.  Its use respects  the context of local quantum field theory, and   allows the  control  of the conformal Ward identities.
 The Beltrami parametrization gives    a formally very strong    parallel between       Yang--Mills   and string theory  BRST  technologies. 
 
One  motivation of this work  is that  string theory is a simpler  arena  than Yang--Mills theory for finding unambiguous  gauge--fixings, beyond the limitation of the Faddeev--Popov method.   
Earlier ideas suggested that, for defining the  Yang--Mills    path integral over $\cal A $/$\cal G $,  in the theory, one should     pick out the   absolute minimum   of  the norm 
 $\int d^4 x \Tr  A_\mu A^\mu $ on each orbit of the space of gauge field configuration  $\{A_\mu\}$. 
 In string theory, one can do 
  a careful and precise analysis.   The result  suggests that one should perhaps use a more refined minimizing function     $\int d^4 x  f( \Tr   A_\mu A^\mu ) $, where the function $f$ is introduced to avoid spurious
   divergencies that do not concentrate at  the boundary of the moduli space  of gauge field configurations.
   In fact, for a given Riemann surface, we will show that one can choose  the following  minimizing function  
    \bea \label{extreme} 
F [\mb, \mbb] =\int_\Sigma  \rho_{z\z} (z,\z)   dzd\bz   \  \frac {1} {{1-|\mu| ^2}}       \ln \frac {1-|\mu|}{1+|\mu|}  .
\eea
where    $\mb$ is the Beltrami differential  and  the  factor $\rho_{z\z}(z,\z) $  is a universal measure that depends only on the genus of the Riemann surface.

 The paper is organized as follows. We first recall   basic formulas for     the Beltrami parametrization in  string theory.  
We  then  explain   the gauge-fixing procedure as a minimizing principle of a relevant functional on the orbit of each Riemann surface.
 We show how it leads  to a BRST-invariant action. The functional expresses the distance between an arbitrarily chosen point in the Teichm\"uller space of a  reference  surface (at a fixed genus) and   any given    possible representative of the 2d-metrics  of a surface, defined modulo local dilatations. The gauge fixing consists in choosing the metric that minimizes this distance. 
The method can be explored in great detail for the torus.  Interestingly, a careful choice of the distance must be done to avoid spurious Gribov-type problems.
For    higher genus, the method is geometrically well-defined, but one  faces  in practice the  complication and/or our  ignorance about  the nature and the details  of    modular groups and fundamantal domains. Technical complications are obviously foreseen for $g>1$, even at $g=2$. 
A formal generalisation of the Nauenberg--Lee--Bloch--Nordsieck arguments to  string theory should be possible in which a cancellation of divergent contributions occurs between amplitudes of different genus, with insertions of ``soft" vertex operators.

Other formulations of string theory  exist where one can find the untwining of the geometry of Riemann surfaces and the quantum field theory of strings, such as the light-cone or the Witten formulation   of open string field theory, see for instance   \cite{zwiebach}\footnote {The idea of such papers is for instance  to show that   the graphs of string theory in light cone quantization are in one-to-one correspondence with Riemann Surfaces, i.e. that each moduli space maps one-to-one into (and onto) a worldsheet diagram.}. However, the conformal gauge approach, as it is formulated in this paper from a minimizing principle, has the great advantage of combining in a rather satisfactorying way a precise description of  Riemann surface orbits and the basic properties of 2-dimensional local quantum field theory.

   \section{Beltrami parametrization  for strings}
   
   \subsection{Definition and the choice of a coordinate system}
   
 Once one understands, following Polyakov \cite{polyakov}, that the propagation of a string on a given manifold sweeps out  quantum mechanically    all possible   worldsheets that can be embedded in a given target space, with possible emissions of other strings,  one needs a parametrization of 2-dimensional manifolds that is as handy as possible, in order to perform a path integral over all the metric  fluctuations. Such a parametrization is provided by the Beltrami differential, which completely avoids the use of the scalar part of the metric,  and provides an appropriate local field variable for the path integral.

  The geometrical data are as follows.  One  considers a metric on an arbitrary smooth compact 2-dimensional Riemann surface 
 $\Sigma (z , \bz)$ 
  without  boundary, and of genus $g$.  Here $(z , \bz)$ denotes at each point a f{i}xed local set of complex analytic coordinates on  $\Sigma $.  The Beltrami differential $\mb$ and its complex conjugate $\mbb$ are   defined by the following parameterization of the 2d-metric on $\Sigma$,
  \be
  ds^2 =\Big(  \exp 2\Phi \Big ) (dz +\mb d\bz )(d\bz +\mbb dz ), 
  \ee
where  $\exp 2\Phi (z , \bz) $ is the conformal factor of the metric in this choice of coordinates. The transformation law of $\mb$ and $\mbb  $ under dilatations  is zero. The  infinitesimal transformations of  $\mb$ and $\mbb  $ will be given shortly in the form of a BSRT symmetry.
  
 A  minimal set of patches  for a   surface  of  a given genus  can be  generally obtained.  The Beltrami differentials are  a set of local functions in each  patch, that      are globally identified on their common boundaries.  When one changes the system of coordinates, $z,\bz \to z' ,{\bz}'$, the shape of the patches changes, but the deformation of their boundaries is obtained by the repametrization in each path, and the identification on the boundaries of neighboring  patches still holds. 
 
 New  coordinates $Z$ and $\bar Z$ are defined by
\bea
dZ = \rho^Z_z (\mb, z,\bar z)(dz+\mb d\bz)    \ \ \ \ \ \ \ d\bar Z =\bar  \rho^{\bar Z }_\bz(\mbb, z,\bar z) (d\bz+\mbb dz).
\eea
 where $ \rho^Z_z$ is an integrating factor. Since  $d^2=0$,  $ \rho^Z_z$  satisfies the differential equation
 $(\partial_{\z} -\mb\partial_z) \ln \rho^Z_z=  \partial _z\mb $
  and  functionally depends only  on  
 $\mb$.

   To define the reparametrization- and dilatation-invariant  quantum field theory that corresponds to a  given  Lagrangian,
   the general approach is to choose once and for all  a fixed set of coordinates.  This means   adopting the point of view of active gauge transformations on the fields, without explicitly changing coordinates.  From now on, we will  thus assume that all fields, including the Beltrami differentials,  only transform under  active symmetries, such as  the BRST symmetry.    All formulas must be written in such a way that they can be  put automatically in correspondence with another system of coordinates. One must  not confuse the BRST symmetry of the theory and the possibility of choosing  different sets of coordinates for defining the path integral. The quantum field theory  is defined as satisfying all Ward identities  corresponding to the BRST symmetry, in the absence of contradictions due to a possible non-vanishing anomaly. Observables are defined from the cohomology of the BRST symmetry. 
     \subsection{2d-action and Beltrami parametrization }

    For any  given local lagrangian  depending    on  the 2d-metric  $g^{\alpha  \beta} $ on   $\Sigma $   and on fields whose arguments are coordinates on $\Sigma $, one can replace the dependence on  $g^{\alpha  \beta} $ by dependence on   $\mb$,   $\mbb$ and  $\Phi$.  For instance, the globally well-defined two-form curvature of $\Sigma $    is 
  \bea
  R_{z,\bz}=  \dz\dzb \Phi +\dz\dz\mb + \dzb\dzb\mbb +{\cal O}(\mu^2)
  \eea
  Conformally-invariant quantities can depend only on $\mb$ and   $\mbb$. Given the string field $X(z,\z)$,  a quick computation  shows that  the Polyakov action is given simply by
  \bea
    \int _ \Sigma  d X \wedge *dX=
  \int _ \Sigma  d^2z \sqrt{g}g^{\alpha  \beta}  \partial _\alpha   X \partial _\beta X
  = \int d^2z \frac { (\dzb - \mb   \dz)X(\dz - \mbb   \dzb)X}{1-\mb\mbb}.
  \eea
For this action,   the path integral over the fields of 2d-gravity   only involves   the Beltrami differential components   $\mb$ and $\mbb$. 
  The gauge-fixing of    Weyl transformations is     trivial, provided that there is no conformal anomaly because, for such conformally-invariant actions, the Faddeev-Popov determinant associated to setting  $\Phi=0$ equals one.  These properties made it possible in the mid-80's to rederive, within the context of local quantum field theory, many conventional results of string theory that had been obtained previously by other methods, e.g.,~\cite{baulieubelloneta}\cite{baulieubelloneta1}. In fact, $\mb$, after its gauge fixing, is nothing but the source of the energy tensor component $T_{zz}$, but to do this gauge-fixing, one must address global issues.  Moreover $\Phi$ is an irrelevant field variable, not seen by conformal invariance.
 
 \subsection{Factorization property and Beltrami parametrization }
   
   The Beltrami parametrization is  well adapted to the factorization property of left and right movers and to the conformal invariance on the  worldsheet that lie at the  heart of string theory.  
   The (active) gauge transformations  of   $\mb$  and $\mbb$
   under    an  infinitesimal  local  2d-diffeomorphism with vector field
  $({\epsilon^z}, {\epsilon^\bz})$     are\footnote{ The relation between $\e$ and the ordinary parameters $\xi$ of diffeomorphism is  $\e^z = \xi^z+\mb\xi^\z$.}
   \def\e{\epsilon^z}
      \def\eb{\epsilon^\bz}
  \bea \label{brsmu}
  \delta \mb &=&\dzb \e   +\e \dz \mb - \mb \dz  \e
  \CR
  \delta \mbb &=&\dz  \eb   + \eb\dzb \mbb - \mbb \dzb  \eb.
  \eea
  We observe that the f{i}elds $\mb(z , \bz)$ and  $\mbb(z , \bz)$  are invariant under local dilatations, and also  that the general infinitesimal  variation of  
  $\mb$ depends only on the single local parameter $\e$, and on    $\mb$.  This is known as the factorization property.
 For the purpose of BRST-invariant quantization,  one 
  replaces  $(\e, \eb)$  by  the anticommuting Faddeev-Popov ghost field $(\cz, \czb)$    and   defines the active  BRST symmetry that corresponds to the  above  infinitesimal transformations  
  \bea \label{brsmu}
  s \mb &=&\dzb \cz   +\cz \dz \mb - \mb \dz  \cz\ \ \ \ \ \ \ \ \ \ s \cz=  \cz\dz\cz
  \CR
  s \mbb &=&\dz  \czb   +\czb \dzb \mbb - \mbb \dzb \czb\ \ \ \ \ \ \ \ \ \ s \czb=  \czb\dzb\czb
  \eea
The action of $s$ is nilpotent on all fields, $s^2=0$.     According to the general  BRST method for local gauge-fixing,   the small diffeomorphism invariance of e.g.\ the Polyakov action can be locally gauged-fixed in the path integral by  adding  to the  invariant classical action an $s$-exact term, which imposes a condition on $\mb$ and  $\mbb$ that allows one to do the path integral.  This gauge-fixing term can be chosen to respect the  left-right independence  on the worldsheet.
However, as in the case of the Yang--Mills theory, no local gauge function can be chosen that is globally well defined; zero modes can occur if one applies the Faddeev--Popov method,  and the way one fixes the gauge for  the 2-d metric must be revisited.

\section{The gauge-fixing question}

Let us now come back to the problem that one faces when  one wishes to sum over all possible Beltrami  differentials $\mb$ and  $\mbb$    
for a given
 Riemann surface $\Sigma$.  Once a set  $\mb$ and  $\mbb$ has been obtained, any other    set  $\{{\mb}^{\cal  G}\} $ and   $\{{\mbb}^{\cal  G}\} $ that is defined     by applying a general diffeomorphism $\cal  G $ on $\mb$ and  $\mbb$ gives another perfectly   equivalent description of the surface.  The space of the Beltrami differential $\mb$  is  connected, but the space of diffeomorphisms is not.  A diffeomorphism  is  either a ``small" one,   composed of a succession of infinitesimal ones, or   a ``large" one, which cannot be connected to the identity transformation, or some combination of such small and large gauge transformations. The    orbit of  $\Sigma$ is therefore a rather complicated  disconnected function in the space of Beltrami differentials, which explains  the difficulty of  the path integral over all   possible Beltrami  differentials.

  The gauge-fixing question is how to  find a way to select a unique representative on each orbit, and how  to make sense of the expectation-value of an observable $\langle \cal O \rangle$ as a well-defined path integral, where the measure of 2d gravity variables only involves the conformal classes of metrics $\mb$ and $\mbb$ : 
 \bea
\langle {\cal O} \rangle = \frac{ \int [d\mb  ] [d\mbb  ] dX   {\cal O}(\mb, X) \exp  \int d^2z \frac { (\dzb - \mb   \dz)X(\dz - \mbb   \dzb)X}{1-\mb\mbb}\ }{  \int [d\mb  ] [d\mbb  ] dX   \exp  \int d^2z \frac { (\dzb - \mb   \dz)X(\dz - \mbb   \dzb)X}{1-\mb\mbb}}\ \ =\ \ \   {\bf \large  ?} 
\eea

  Non-perturbatively, the conventional Faddeev--Popov method   generally   fails,   as explained   very clearly by  Singer in the Yang--Mills theory, since  the so-called gauge condition is in fact not globally well-defined~\cite{gribov}. In the present  case the   local gauge condition cuts    orbits erratically, and all sorts of inconsistencies may occur.  For example, the conformal gauge  consisting in taking   $\mb=\mbb=0$ can only be imposed locally, otherwise it selects only the square torus.  

As compared to the Yang--Mills case, the difficulty that occurs in the conformal gauge for 2d gravity  is  analogous to the so-called Gribov ambiguity of a Landau-type gauge. It is  however much   simpler to handle, and even to describe, because in the case of 2d-gravity we have a good understanding  of  the orbits of Beltrami differentials.

An advantage of the string situation is that, from the beginning,  we deal with bounded functions. One has the constraint that any representative on the orbit must  satisfy  
  $\det\sqrt {g}= \exp \Phi  |{1-\mb\mbb}|$ cannot vanish. In fact, on any given point of an orbit,  positivity requires
 \bea
 | \mb |  <1
  \eea

In the case of the torus, this property  justifies the  use the  Poincar\'e disk  $ \cal D$ of complex numbers $|\gamma| \leq 1$,   as a representation of  the Teichm\"uller space, instead of the upper-half space $Im (\tau) \geq 0$. One  foresees that any complication, if it occurs, can only happen for the singular points of the  boundary of the moduli space, which is the unit circle in the case of the torus.  However,   already in this  simple case the use of the conformal gauge, treating $\mb$ and $\mbb$ independently,  is too naive, and global questions must be  addressed {\it ab initio.} 

The so-called conformal gauge, in which  one  tries to 
gauge-f{i}x $\mb$ and  $\mbb$ to a given background with $\Phi=0$,  is not     compatible with the global structure of   $\Sigma (z , \bz)$. Taking $\Phi=\mb=\mbb=0$ is a   much  too strong  condition since it   implies that $  R_{z,\bz}=0 $    everywhere, which is generally wrong, and   the brute force application  of the perturbative BRST method for the conformal gauge  explicitely leads to inconstancies, under the form of zero modes for the Faddeev--Popov operator. Even if the problem can be corrected by trial and error, eventually   giving  a partition function that reduces to an integral over a fundamental domain (when the modular group is known), or over the Teichm\"uller space (modulo some denumerable redundancy),  logically one should not start the process by gauge-fixing in the conformal gauge.

In the case $g=1$,  among all equivalent representations of a torus in   $ \cal D$, we will give a criterion for  choosing  a unique representative. For higher genus, $g>1$, the problem is more intricate, but our approach still holds, and we will explain it first, and then check the consequences for the torus. 


%

\section{Choosing a minimizing gauge-functional to define the 2d-gravity path integral.}

%
 
 To    select a unique representative for  the Beltrami differential, we   propose a minimizing functional, $F[\mb, \mbb]$ to be extremised, orbit by orbit, in the space of Beltrami differentials.  This functional represents a possible distance between the Beltrami parametrization $\mu$ of any given Riemann surface $\Sigma$ and that of an arbitrarily chosen Riemann surface of the same genus, whose representative is also freely chosen on its gauge orbit.\footnote{The functional, $F[\mb, \mbb]$,  and the  ``distance" it represents, will be used to gauge fix, that is to say, to select one representative out of all possible gauge-equivalent configurations, so naturally it will not itself be gauge invariant.}  We denote by $\Gamma$ the chosen  representative of  the   Beltrami differential of this reference surface.  One must check eventually that $\Gamma$ can be changed without affecting the values of observables, a property that can be demonstrated by  the Ward identities of the underlying BRST symmetry of the construction.

    The minimizing process must be done in several steps. One starts from a given point $\mb$ on the orbit of $\Sigma$, and minimizes the functional $F$ with respect to gauge transformations along the orbit that are connected to the identity, and gets  a point in the Teichm\"uller  space.  Then one looks  for all other  extrema that are connected to the former one  by large gauge transformations, that is, the diffeomorphisms that are not connected to the identity, and gets down to the moduli space. The gauge-fixing  consists in finding the absolute minimum among all these local extrema.

We choose  to express the distance from $\mb$ to $\Gamma$ by
\bea \label{extreme1} 
F^\Gamma[\mb, \mbb] =\int_\Sigma    dzd\bz   \  \rho_{z\z}(z,\z)  \ \frac {D(\mb, \Gamma, \mbb, \bar \Gamma) }{1-\mb\mbb }.
\eea
The   factor $\rho_{z\z}(z,\z) $  is a measure that exists  for any given set of coordinate patches $\{z,\z\}$, and allows one to make  the integral~(\ref{extreme1}) well defined. It is a universal factor that is the same for all surfaces of given genus $g$.  Consequently 
when $\mb$ runs along an orbit, $\rho_{z\z}(z,\z) $ remains the same.  (For the torus, that is $g=1$, one can chose $\rho_{z\z}(z,\z)=1 $.)  Therefore, when we 
   look for a local    extremum of $F[\mb, \mbb]$ under transformations that are continuously connected to the identity, we will vary $\mb, \mbb$, while keeping $\rho_{z\z}(z,\z) $ as a fixed measure for all surfaces of the same genus.


The motivation for the factor $(1-\mb\mbb )^{-1}$ that diverges at $|\mb|\sim1$ is as follows. In the case of the torus, we found that this factor allows one to concentrate all possible ambiguities at  the singular point of the boundary of the Teichm\"uller space. In fact we shall show that, with this factor, the value of $\mb$ that extremises the variation of $\mb$ under the action of small diffeomorphisms is an absolute minimum rather than a saddle point.  (Relative minima that are not absolute do not occur). This allows one for instance to understand the gauge-fixing as resulting from a drift force that is always attractive, everywhere on the orbit. 

We consider explicitly the case where one may choose $\Gamma=0$, and 
\bea
D(\mb, \Gamma, \mbb, \bar \Gamma)= D(\mb \mbb).
\eea
  With further knowledge of the theory of Riemann surfaces,  when $\mb$ is identified with a representative  of the Teichm\"uller space, the function $D$ can be understood as a possible distance in this space. The functional  $F$  is the lift of this  distance   in the space of  the conformal classes of metrics $\mb$ and $\mbb$, by the inverse operation of the`` small"  diffeomorphisms $ {\cal D}iff_0$ that are connected to the identity.

To simplify notation, we now define 
\bea\label{minmin}
f(x) = \frac {D(x)}{1-x}
\eea
so that 
\bea \label{extremea} 
F [\mb, \mbb] =\int_\Sigma    dzd\bz   \  \rho_{z\z}(z,\z)  f ( \mb\mbb ) 
\eea
Having in mind the  relevance of the so-called Weil-Petersen metric, we can propose
\bea 
\label{min2}
D(x) = \ln \frac {1-x}{1+x}
\eea
or 
\be \label{min1}
D(x) = \ln \frac {1-\sqrt x}{1+\sqrt x}.
\ee
One may prefer that the distance between 2 points be linear in $|\mb|$ for small $|\mb|$, in which case the second choice is preferable. For the sake of the minimization principle for $F[\mb, \mbb]$  along a gauge orbit, we will check that both choices (\ref{min2}) and  (\ref{min1}) are acceptable, and a wider class of $f(x)$ may also be considered.

\section{Extremals of  the gauge function}

\subsection{ Extremisation equation for $\mb$ and its resolution}

When the functional (\ref {extremea}) is at a  local extremum under infinitesimal coordinate transformations, the stationarity condition is  
\bea
\label{firstdelta}
\delta F(|| \mu ||^2 )& = & \int dz d\bar z    \ \rho_{z\z}(z,\z)    \ (  \mbb \delta \mb + c.c. )f'  
\nonumber \\
& = & \int dz d\bar z    \ \rho_{z\z}(z,\z)    \   \mbb ( \dzb \e   +\e \dz \mb - \mb \dz  \e ) f' + c.c.
\nonumber \\
& = & - \int dz d\bar z \    \  \epsilon^z (  \p_\z - \mb \p_z  - 2   \p_z \mb) 
(\rho_{z\z} \mbb f') + c.c. 
\eea
It is convenient to introduce the tensor with componants
\be
h_{zz}(\mb, \mbb, z, \z) \equiv \rho_{z\z}(z, \z) f'(\mb \mbb)  \mbb,
\ee
and c.c., and a local extremum is characterized by the equations,
\bea
 \label{extremumcond}
(\dzb -     \mb   \dz -   2  \dz\mb) h_{zz}  = 0,
\eea
and c.c.

 For the torus,  $g=1$, one can take $   \rho_{z\z}(z,\z) =1$, so $h_{zz} =  f'(\mb \mbb)  \mbb$ has no explicit dependence on $z$ and $\bar z$.  In this case the solution of \eqref{extremumcond} for $\mb$ is
\be
\mb = \gamma,
\ee
where $\gamma$ is a constant (complex) modulus, defined modulo an $SL(2,Z)$ transformation. 
 
    For genus $g > 1$,  one uses the Riemann-Roch theorem to solve \eqref{extremumcond}.  One goes to another system of coordinates $Z,\bar Z$, such that 
$
  dZ =\Lambda^Z_z (dz +\mb d\z) $ and $ d\bar Z =\Lambda^{\bar Z }_z (d\z +\mbb dz) 
$. As noted earlier,  the integrating factor, $\Lambda^Z_z $, depends functionally only on $\mb$,  
$ \Lambda^Z_z =  \Lambda^Z_z (\mb,  z,\z) $ and $\Lambda^{\bar Z }_z =  \Lambda^{\bar Z }_z (\mbb,  z,\z) 
$. Then the 
 equation,
  \bea
  \label{Z}
  \partial _{\overline Z} H_{Z,Z} =0,
  \eea
   implies by the Riemann-Roch theorem that 
 \bea H_{Z,Z} =\sum _{1\leq i\leq 3g-3} \gamma^i H_i(Z).\eea
  The $\gamma^i$ are complex moduli that can be chosen to vary over any given fundamental domain. The
   $H_i(Z)$ are a  basis of  the 3g-3 zero modes of quadratic differentials that is to say, the 3g-3 linearly independent (complex) solutions of \eqref{Z}. \def\Z{{\bar Z}}  Now   by tensorial covariance, one has 
\bea
 h_{zz}= (\Lambda^Z_z)^2   H_{Z,Z}= \sum _{1\leq i\leq 3g-3} \gamma ^i h_i   (\mb,z,\z),
\eea
which satisfies \eqref{extremumcond}.

     The solution of the  minimizing equations is thus given by   
    \bea
\rho_{z\z}(z, \z) f'(\mb \mbb)  \mbb    =  \sum _{1\leq i\leq 3g-3}  \gamma^i h_i(\mb, z,\z)\nonumber
\eea 
      \bea\label{sol}
\rho_{z\z}(z, \z) f'(\mb \mbb)  \mb   =  \sum _{1\leq i\leq 3g-3}  \bar \gamma^i  \overline h_i(\mbb, \z,z).
  \eea
 These are a pair of coupled functional equations for $\mb$ and $\mbb$ with solutions
 \be
 {\mb} = {\mb}_0(\gamma, \bar\gamma, z,\z)
 \ee
 and c.c. We emphasize that here the dependence on the $\gamma$'s is highly non-linear, and it is a challenge to find the solution explicitly even for $g = 2$.

 Let us summarize the situation. Equation \eqref{extremumcond}, which determines an extremum of the functional (\ref{extremea}), is solved when  $h_{zz}$ is expressed as a linear  combination of  particular functions $h_i$, with complex coefficients $\gamma^i$.  The $\gamma^i$  can be identified as a point in the Teichm\"uller space.
Thus, starting from  an arbitrary point $\mb$ on  the orbit, one can reach a point of the Teichm\"uller space by a succession of small gauge transformations that brings one to a stationary point on the gauge orbit which is a minimum with respect to all small gauge transformations.  The modular  group, which consists of the large gauge transformations, allows one to jump discontinuously from one stationary point to any other stationary point on the orbit.  By choosing the absolute minimum of $F$ among the stationary points on each orbit, we obtain a fundamental modular region that contains the reference point $\gamma = 0$, and provides a unique representative for each Riemann surface (modulo local dilatations).




 \subsection{ Behaviour of the orbit near the local extremum ${\mb}_0$}
  Because eqs.~\eqref{sol} depend on $f$, the solutions ${\mb}_0$  depend implicitly on the choice of $f$. 
    In order to obtain only minima of the minimizing functional $F$ rather than extrema that are merely saddle points, one may try to  choose  the function $f$ such that the matrix of second derivatives of the minimizing functional is always positive in the fundamental modular region (except at singular points that occur on the boundary of the fundamental domain, and correspond to degenerate Riemann surfaces, such as the pinched torus).  This property ensures that when one applies a small diffeomorphism to ${\mb}_0$, so that the representative of the surfaces exits the Teichm\"uller space, its norm can only grow. If this property can be ensured throughout a fundamental domain,  one gets a Hessian that is positive definite everywhere (but at the singular point(s) of the  fundamental domain).  We will show (in the case of the torus) that it permits one to describe the gauge fixing as the result of an attractive drift force along the orbit  via stochastic quantization.  The criterion is that the behavour of  $f(x)$ is sufficiently near the  horizon, at $x=1$.
    
\begin{figure}
\begin{center}
\includegraphics[width=14cm]{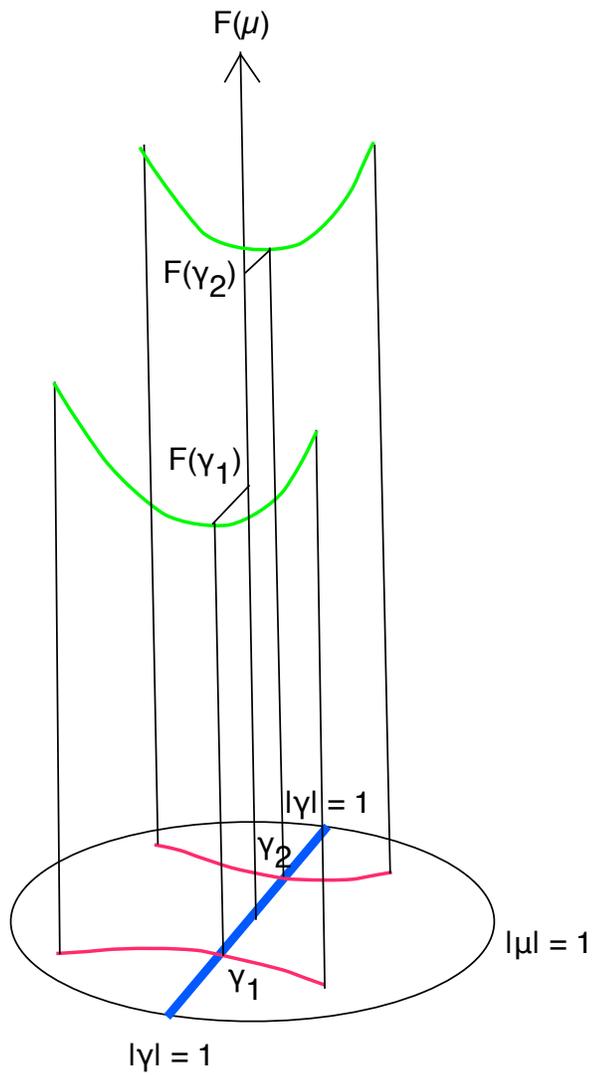}
\caption{{Plot of the values of the minimizing functional, in green, corresponding to a single gauge orbit, in red, of which only two (out of an infinite number of) disconnected branches are shown.  The blue horizontal line is the Teichm\"uller space.}}\label{functionalorbit}
\end{center}
\end{figure}
\hspace{-0.6cm}
    This situation is pictured in Fig.~\ref{functionalorbit}.  The infinite-dimensional space of the $\mu(z, \bar z)$ is represented in perspective in the horizontal plane.  It contains the Teichm\"uller space represented by the horizontal blue line segment.  A single gauge orbit, consists of an infinite number of disconnected branches, of which only two are shown in the figure.  They are represented by the two disconnected horizontal red curves that intersect the Teichm\"uller space at $\mu = \gamma_1$ and $\mu = \gamma_2$.  The Teichm\"uller parameters $\gamma_!$ and $\gamma_2$ are related by a `large' gauge transformation $\in SL(2, Z)$.  Each red curve is related to $\gamma_1$ or $\gamma_2$ by a `small' gauge transformation that is continuously connected to the identity.  The vertical axis measures values of the minimizing functional $F(\mu)$, and the two green curves show the values of $F(\mu)$ for points $\mu$ on the gauge orbit, just described, that is obtained from the green curves by vertical projection.  The green curves are at a minimum at $F(\gamma_1)$ and $F(\gamma_2)$, where the branches of the gauge orbit intersect the Teichm\"uller space.  
    
    An interesting feature is that there can be only a single minimum of the minimizing functional on each connected branch of a gauge orbit.  Indeed, suppose that there were two relative minima on the same branch.  In this case they are related by a gauge transformation that is continuously connected to the identity.  On the other hand each minimum satisfies the stationarity condition, which means that each minimum is a point in the Teichm\"uller space.  However, within the Teichm\"uller space, two points that are gauge-equivalent are related by a  large diffeomorphism, which cannot be continuously connected to the identity.  Thus we have arrived at a contradiction, which shows that there cam be only a single minimum on each connected branch of a gauge orbit.  We shall show explicitly for the case of the torus that the single minimum does in fact exist for appropriately chosen minimalising functional.

 
\section{BRST-invariant action}

    We would like to impose the above gauge fixing in a BRST-invariant way.  For this purpose, we introduce the gauge-fixing  BRST-exact Lagrangian,
\bea
 s \int dzd\z 
  [b_{zz} (\mb -{\mb}_0) + c.c.] & = & \int dzd\z \Big( \lambda_{zz} (\mb -{\mbb}_0) - b_{zz} (\dzb \cz   +\cz \dz \mb - \mb \dz  \cz) \Big)  \nonumber  \\
  & + &\sum _{1\leq i\leq 3g-3}  l^i  \int dzd\z       \frac {\partial  {\mb}_0 (\gamma,\bar\gamma ,z,\z)    }{ \partial  \gamma^i}b_{zz} + c.c.,
   \eea
 where the BRST-operator $s$ acts according to
 \bea \label{brsmu1}
  s \mb &=&\dzb \cz   +\cz \dz \mb - \mb \dz  \cz\ \ \ \ \ \ \ \ \ \ s \cz=  \cz\dz\cz
  \CR
   s b_{zz}  &=& \lambda_{zz}\ \ \ \ \ \ \ \ \ \ \ \ \ \ \ \ \ \ \ \ \ \ \ \ \ \ \ \ \ \ \ \ \ \ \ \ \  s  \lambda_{zz} =  0
     \CR
   s \gamma^i  &=& l^i\ \ \ \ \ \ \ \ \ \ \ \ \ \ \ \ \ \ \ \ \ \ \ \ \ \ \ \ \ \ \ \ \ \ \ \ \ \ \ \ \   s  l^i=  0,
  \eea
and c.c., with $s^2 = 0$.  Here the $\gamma$ are in a fundamental  domain containing the value $\gamma=0$. The Lagrange multiplier $\lambda_{zz}$ assures that the minimization condition on $\mb$ is satisfied on each gauge orbit, and this value of $\mb$ automatically gets substituted everywhere in the action and the observables.
    
    The last term in the action imposes, by integration over the $l^i$, that the antighost field  $b$ remains orthogonal to all zero modes in the  Fadeev--Popov operator, defined by   $  (\dzb    -  \mb \dz+  2  \dz \mb)      b_{zz}  =0$.  
    
    We will check  that no zero eigenvalue occurs for the torus, by an appropriate choice of the function $f$. In fact  the zero mode occurs only at the singular part of the boundary of the minimizing  fundamental domain, which constitutes therefore a harmless Gribov horizon.

The definition  of  the observables as $s$-invariant quantities that are not $s$-exact ensures that they cannot depend on the $\gamma$'s, because the pairs $(\gamma, l)$  are BRST-trivial doublets.   Their field dependance is only   though   the string field $X$ and the Beltrami differentials $\mb$ and $\mbb$ (and their supersymmetric partners in the superstring case).

The alternative of  imposing   equations \eqref{extremumcond} as gauge conditions, by means of Lagrange multiplier fields  in a standard BRST-invariant way will be  sketched in an  Appendix.  However this gauge choice is impractical because gravitational degrees of freedom propagate, and for this reason we shall impose instead the solution of this equation, which is $\mu_{\overline z}^z = \gamma$ in the case of the torus.

 \section{The case of the torus}
 
 \subsection {Identification of the domain that minimizes the gauge-functional}

 For the torus, the Teichm\"uller space can be represented as the upper half-plane of complex   $\tau$, with 
  $Im( \tau)\geq  0 $. Two points that differ by any given $SL(2,Z)$ transformation
  \bea
  \tau \to \frac {a\tau +b}{c\tau +d},
  \eea
   where $a,b,c$ and $d$ are positive or negative integers, represents the same torus. These transformations can be decomposed as successions of transformations
   \bea\label{mapping}
  \tau \to \frac {-1 }{ \tau  }; \ \ \ \ \ \ \  \tau \to  { \tau +1 }.
  \eea
 As is well known, the first fundamental domain is defined by     $  -\frac {1}{2}\leq    Re (\tau)   \leq  \frac {1}{2}$
 and $\tau \bar \tau \geq 1$. All other fundamental domains are obtained from compositions of transformations~(\ref{mapping}).
 
%

For any given Riemann surface one has everywhere   
$|\mb|  <1 $.     It  is thus appropriate to redefine the Teichm\"uller parameters in such a way that they are confined in a disk where their modulus remains smaller than one. For the torus, the solution is obvious; all points of the complex upper-half plane {\cal{ Im}}$\  \tau \geq 0$ are mapped
onto the Poincar\'e disk  $\cal D$, $|\gamma| \leq 1$, by
  \bea
  \gamma= \frac{ \tau -i}{\tau +i}
  \eea
  As we will show, this opens the way to the gauge-fixing of 2d-gravity in a very simple way.

  Figure~2 shows the image    $\cal D_I$  of the first fundamental domain for the case of the torus. The mapping   $  \tau \to \frac {-1 }{ \tau  }$ corresponds to a symmetry $  \gamma \to - \gamma $ for  every point of any given fundamental domain. The mapping $ \tau \to  { \tau +n }$ sends a     point  $\gamma$ of the domain $I$ into a point  
  $\gamma_n $ of another fundamental domain such that  $ \gamma_n  >  \gamma  $. 

      The curves in $\gamma$-space that appear in the figure are found from the inversion,
\bea
\tau = i \Big( {1 + \gamma \over 1 - \gamma } \Big) = i \Big( {1 + x + iy \over 1 - x - iy } \Big),
\eea
where we have separated $\gamma$ into its real and imaginary parts.  One easily finds that the boundary of the Teichm\"uller space, ${ Im} (\tau )= 0$ corresponds in the $\gamma$ plane to the unit circle $x^2 + y^2 = 1$, whereas the boundary of the first fundamental modular region, made up of parts of the curves ${  Re} (\tau) = \pm 1/2$, and $|\tau| = 1$, is made up in the $\gamma$ plane by parts of the curves $(x-1)^2 + (y \mp 2)^2 = 2^2$ and $x = 0$, as drawn in Fig.~2.
   \begin{figure}
\includegraphics[width=12cm]{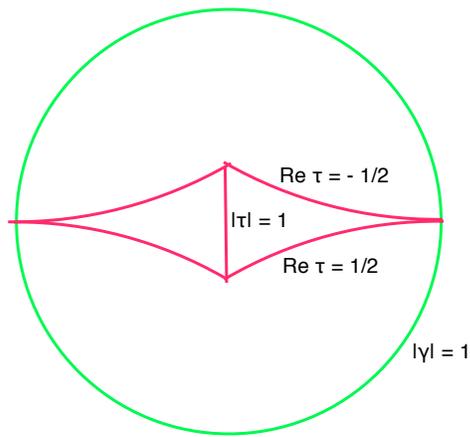}
\caption{ {The  Poincar\'e disk $|\gamma| \leq1$  corresponds to the Teichm\"uller space ${\rm Im} \tau \geq 0$.} The first fundamental modular region of the torus, and its copy under $\gamma \to - \gamma$, are outlined in red, as  the interior of both adjacent triangles in the  middle  of the $\gamma$ plane. For these domains, the point   $|\gamma| =0 $  is the     representative of the ``squared" torus.  This point can be e.g. chosen as the reference point for the minimizing functional in string theory. All other fundamental domains are obtained by modular transformations. Each one of them intersects only once the boundary circle $|\gamma| =1 $. On the other hand any given point of this boundary  belongs  to several fundamental  domains. The  boundary of the  Poincar\'e disk can be therefore  named as the (harmless)   horizon of 2-dimensional gravity.}
\label{FMR}
\end{figure}
\hspace{-0.6cm}
  
%
%

  \newpage
   \subsection {BRST-invariant representation of the   minimizing domain}
 For genus $g = 1$, the last Lagrangian simplifies to
  \bea\label{brstgf}
  s( \int dz d\bar z \Big (   b_{zz}(\mb -\gamma) \Big) =
  \int dz d\bar z  \Big (   \lambda_{zz}(\mb -\gamma)     
  -    b_{zz}   (\dzb \cz    - \gamma  \dz  \cz) \Big)    +l  \int dz d\bar z\  b_{zz}
  \CR+ c.c.,\ \ \ \ \ \ \ \ \ \ \ \ \ \ \ \ \  \ \ \ \ \ \ \ \ \ \ \ \ \ \ \ \ \ \  \eea
where $s$ acts as in \eqref{brsmu1}.  
 This expression must be added to the Polyakov action, which is BRST-invariant but not BRST-exact.    The  gauge-fixing action~(\ref{brstgf}) identifies   $\lambda_{zz} $, as a Lagrange multiplier field for $\mb$.  The constant fermionic Lagrange multiplier $l$ imposes that the zero mode of the operator $\dzb      - \gamma \p_z  $ is omitted. Consequently the ghost and anti-ghost integrations give a regularised determinant,   $\det' ( \dzb      - \gamma )$.
 Eventually, the integration over $\gamma$ must be done over the fundamental domain that we found by our minimizing principal for each orbit.  This reproduces the known result for the partition function of string theory with a 1-torus worldsheet. In this construction, 
it must be noted that, although one has escaped the consequence of Singer's theorem by  solving a   minimizing principle,  a BRST symmetry has been preserved all along the way, allowing one to prove by locality properties that the observables  satisfy all requirements concerning factorization and modular invariance.
Notice that the degenerate point $\gamma=1$ is safely approached.  This is where the torus approaches the pinched torus, that is, a sphere with two identified points.  If an observable produces divergences as one approaches this point, one must e.\ g.\ use a cutoff $| \gamma| < 1-\epsilon$, consistent with the BRST Ward identity (see the previous section), so the divergence cancels in the limit $\epsilon\to 0 $. 

We now verify the absence of zero modes of the second variation of the minimizing functional, except at the singular point $|\gamma|=1$, and we determine the criteria on the function $f$ in order that the second variation of the minimizing functional $F$ be strictly positive for  $\gamma <1$.

  \subsubsection{   Eigenvalues and zero modes of the Faddeev-Popov operator } 

    We shall calculate the eigenvalues of the Faddeev-Popov operator

    \be
\label{FP}
M \equiv \overline{\p} - \gamma \p  
\ee     
where $\p \equiv {\p \over \p z}$ and  $\overline\p \equiv {\p \over \p \overline z}$. \

  This operator acts on functions $f(z, \overline z)$ that are doubly periodic in the basic parallelogram
\be
\label{first}
f(x + 1, y) = f(x, y + 1) = f(x, y).      
\ee
where $z = x + i y$, and $x$ and $y$ are real.  Note that the boundary conditions satisfied by the coordinates are fixed, independent of the metric, because our transformations are all active, that is to say, they act on the fields only. 

The Faddeev-Popov operator is a derivative with constant coefficients which is diagonalized by an exponential,
\be
f_{m, n}(x, y) = \exp[2 \pi i (m x + n y)], 
\ee
and the boundary conditions are satisfied by taking $m$ and $n$ to be integers.  Thus the general solution with the doubly periodic boundary conditions reads, in terms of $z$ and $\overline z$
\be
f_{m, n}(x, y) = \exp [ \pi (mi + n ) z + \pi (mi - n) \overline z ].
\ee

The eigenvalues of the Faddeev-Popov operator are obtained from 
\be
(\overline{\p} - \gamma \p) f_{m, n} = E_{m,n} f_{m, n},
\ee
which gives
\be
E_{m, n} = \pi (mi - n) - \gamma \pi (mi + n).
\ee
The null eigenvalues satisfy $E_{m, n} = 0$, which gives for the values of $\gamma$ that correspond to null eigenvalues,
\be
\label{zeromodes}
\gamma = - { n - mi \over n + mi }.
\ee
This implies
\be
| \gamma | = 1,
\ee
and so all values of $\gamma$ that correspond to zero-modes of the Faddeev-Popov operator $M = \overline{\p} - \gamma \p$ lie on the unit circle.

\subsubsection{Second  variation of minimizing functional}

\def\pa{\partial}
 
 The derivation of the second variation of $F$ is simplified by never partially integrating on $\mu$ or $\bar\mu$ because, in the end, the condition of minimisation that is imposed, is $\mb=\gamma$ = const. For notational simplicity we now set $\mu=\mb$ and $\bar \mu=\mbb$.
 
 The minimizing functional is given by
 \be
 F= \int d^2z \ \rho \ f( \bar\mu \mu).
 \ee
 Its first variation is
 \bea
 \delta F& = & \int d^2z \ \rho \ f'( \bar\mu \mu) \ (\delta \bar\mu \mu + \bar\mu \delta \mu)  \nonumber \\
 & = & \int  d^2z \ \rho \ f'( \bar\mu \mu) \ (\nabla \bar\epsilon \mu + \bar\mu \overline\nabla \epsilon),
 \eea
 where we have used
 \be
 \label{deltamu}
 \delta \mu = \overline\nabla \epsilon \equiv (\bar\pa - \mu \pa + \pa \mu) \epsilon, 
 \ee
 and cc.  
 
 The second variation is then
 \bea
 \delta^2 F& = & \int d^2z \ \rho \ \{ \ f''( \bar\mu \mu) \ (\nabla \bar\epsilon \mu + \bar\mu \overline\nabla \epsilon)^2  \nonumber \\
 &  & \ \ \ \ \ \ \ \ \ \ \  + \  f'( \bar\mu \mu) \ [ \ 2 \nabla \bar\epsilon \overline\nabla \epsilon + \delta ( \nabla \bar\epsilon ) \mu + \bar\mu \delta ( \overline\nabla \epsilon) \ ] \   \}.
 \eea
 By \eqref{deltamu} we have $\delta \overline\nabla \epsilon = (- \delta\mu \pa + \pa \delta\mu) \epsilon$, which gives
 \be
\delta \overline\nabla \epsilon = - \overline\nabla \epsilon \pa \epsilon + \pa (\overline\nabla\epsilon)  \epsilon
 \ee
 and cc, and we have
 \bea
 \delta^2 F& = & \int d^2z \ \rho \ \Big[ \ f''( \bar\mu \mu) \ (\nabla \bar\epsilon \mu + \bar\mu \overline\nabla \epsilon)^2   \\  \nonumber
 &  & \ \ \ \ \ \ \ \ \ \ \  + \  f'( \bar\mu \mu) \ \Big( \ 2 \nabla \bar\epsilon \overline\nabla \epsilon + [- \nabla \bar\epsilon \bar\pa \bar\epsilon + \bar\pa (\nabla \bar\epsilon) \bar\epsilon ] \mu + \bar\mu [- \overline\nabla \epsilon \pa \epsilon + \pa (\overline\nabla\epsilon) \epsilon ] \ \Big) \   \Big].
 \eea
 
 We are interested in the second variation at the stationary points of the minimizing functional, and we specialize to the torus.  In this case we have $\rho = 1$, and $\mu = \gamma$ and $\bar\mu = \bar\gamma$, where $\gamma$ and $\bar\gamma$ are complex conjugate constants with $\gamma \bar\gamma \leq 1$.  In this case $\overline\nabla \epsilon$ simplifies to
 \be
 \label{simplifies}
\overline\nabla \epsilon \equiv (\bar\pa - \gamma \pa) \epsilon, 
 \ee
 and cc, and we have
 \bea
 \delta^2 F|_{\mu = \gamma} & = & \int d^2z \ \Big[ \ f''( \bar\gamma \gamma) \ (\nabla \bar\epsilon \gamma + \bar\gamma \overline\nabla \epsilon)^2   \\  \nonumber
 &  & \ \ \ \ \ \ \ \ \ \ \  + \  f'( \bar\gamma \gamma) \ \Big( \ 2 \nabla \bar\epsilon \overline\nabla \epsilon + [- \nabla \bar\epsilon \bar\pa \bar\epsilon + \bar\pa (\nabla \bar\epsilon) \bar\epsilon ] \gamma + \bar\gamma [- \overline\nabla \epsilon \pa \epsilon + \pa (\overline\nabla\epsilon) \epsilon ] \ \Big) \   \Big].
 \eea
 We simplify this expression by doing a partial integration in the last two terms,
 \bea
 \delta^2 F|_{\mu = \gamma} & = & \int d^2z \ [ \ f''( \bar\gamma \gamma) \ (\nabla \bar\epsilon \gamma + \bar\gamma \overline\nabla \epsilon)^2  
 \nonumber   \\ 
 &  & \ \ \ \ \ \ \ \ \ \ \  + \ 2 f'( \bar\gamma \gamma) \ (  \nabla \bar\epsilon \overline\nabla \epsilon - \nabla \bar\epsilon \bar\pa \bar\epsilon \gamma - \bar\gamma  \overline\nabla \epsilon \pa \epsilon ) \  ].
 \eea
 
 \subsubsection{Positivity of eigenvalues}
 
 We wish to determine if the second variation, $\delta^2 F|_{\mu = \gamma}$, is a positive quadratic form.  Since it involves derivatives with constant coefficients, we may diagonalize it by fourier components.  With coordinates $z = x + iy$ and $\bar z = x - iy$, the boundary conditions for the torus are
 \be
 \epsilon(x + 1, y) = \epsilon(x, y+1) = \epsilon(x, y).
 \ee
 and cc.  The second variation is diagonalized by
 \be
 \epsilon(x, y) = \alpha \sin[2 \pi (mx + ny)] + \beta 
\cos[2 \pi (mx + ny)],
 \ee
 and cc, where $m$ and $n$ are integers, and $\alpha$ and $\beta$ are complex constants.  Since $\delta^2 F|_{\mu = \gamma}$ is quadratic in the derivatives $\pa \epsilon, \pa\bar\epsilon, \bar\pa\epsilon, \bar\pa\bar\epsilon$, the terms in $\sin[2 \pi (mx + ny)]$ and $\cos[ 2 \pi (mx + ny)]$, do not mix, so the terms in $\alpha$ and $\beta$ do not mix, and we may diagonalize by taking $\beta = 0$ or $\alpha = 0$.  These choices give the same result, and we take
 \bea
 \epsilon(x, y) & = & \alpha \sin[2 \pi (mx + ny)]
 \nonumber  \\
 \bar \epsilon(x, y) & = & \bar\alpha \sin[2 \pi (mx + ny)]
 \eea
 We have 
 \bea
 \pa = \pa_z = (1/2)(\pa_x -i \pa_y)
 \nonumber \\
 \bar\pa = \pa_{\bar z} = (1/2)(\pa_x +i \pa_y),
 \eea
 which gives
 \bea
 \pa \epsilon = \overline W \alpha \cos[2 \pi (mx + ny)]; \ \ \ \ \pa \bar\epsilon = \overline W \bar\alpha \cos[2 \pi (mx + ny)]  
 \nonumber   \\
 \bar\pa \epsilon = W \alpha \cos[2 \pi (mx + ny)]; \ \ \ \ \bar\pa \bar\epsilon = W \bar\alpha \cos[2 \pi (mx + ny)]
 \eea
 and
  \bea
 \overline \nabla \epsilon = V \alpha \cos[2 \pi (mx + ny)]
 \nonumber  \\
  \nabla \bar\epsilon = \overline V \bar\alpha \cos[2 \pi (mx + ny)],
 \eea
 where
 \be
 W \equiv \pi(m + in); \ \ \ \ \ \  \overline W \equiv \pi(m-im).
 \ee
 and
 \bea
 \label{VandbarV}
 V & \equiv & W - \gamma \overline W \nonumber  \\
 \overline V & \equiv & \overline W - \bar\gamma W.
 \eea
 
 Upon integrating over $x$ and $y$, we obtain for the second variation
 \bea
 2 \delta^2 F|_{\mu = \gamma} = f'' ( \overline V \gamma \bar\alpha +  V \bar\gamma \alpha)^2
 + 2 f' ( |V|^2 \bar\alpha \alpha 
 - V \overline W \bar\gamma \alpha^2
 - \overline V W \gamma \bar\alpha^2 ).
 \eea
 In terms of the variables $\alpha$ and $\bar\alpha$, this is the quadratic form
 \be
 \label{form}
2 \delta^2 F|_{\mu = \gamma} = A \bar\alpha \alpha + B \alpha^2 + \overline B \bar\alpha^2,
 \ee
 where
 \be
 A \equiv 2 f'' |V|^2 \bar\gamma \gamma + 2 f' |V|^2
 \ee
 \be
 B \equiv f'' V^2 \bar\gamma^2 - 2 f' V \overline W \bar\gamma,
 \ee
 and cc. 
    In terms of the real variables
 \be
 \alpha = r + is; \ \ \ \ \ \bar\alpha = r - is
 \ee
 it reads
 \be
 \label{realform}
2 \delta^2 F|_{\mu = \gamma} = (A + B + \bar B) r^2 + (A - B - \bar B) s^2 + 2i(B - \bar B) rs.
 \ee
 The eigenvalues of this real quadratic form are easily found to be
 \be
 \lambda = A \pm 2 \bar B B.
 \ee
 For appropriately chosen $f(x)$, the derivatives $f'(x)$ and $f''(x)$ are positive, so $A$ is positive, and both roots will be positive if $A^2 > 4 \bar B B$ namely, if
 \be
 A^2 - 4 \bar B B > 0.
 \ee
 We wish to determine if this quantity is positive for all values of $W$ and $\gamma$, with $\bar\gamma \gamma \leq 1$.  
 
 	To simplify the calculation we write
 \be
V = W v; \ \ \ \ \ \overline V = \overline W \bar v
\ee
where, by \eqref{VandbarV},
\be
v \equiv 1 - \sigma; \ \ \ \ \ \ \bar v = 1 - \bar \sigma
 \ee
 and
 \be
 \sigma \equiv \gamma {\overline W \over W} = |\gamma| e^{i\phi}; \ \ \ \ \ \ \bar\sigma \equiv \bar\gamma {W \over \overline W} = |\gamma|e^{-i\phi}.
 \ee
Here $\phi$ is a pure phase because
\be
 {\overline W \over W} = {m - i n \over m + in },
 \ee
is a pure phase,
and we have
\be
\bar\sigma \sigma = \bar\gamma \gamma.
\ee
In terms of these variables we have
\bea
A & = & 2 f' |V| |W| |v| (R |\gamma|^2 + 1)
\nonumber   \\
B & = & f' V \overline W \bar \gamma (R v \bar\sigma -2)
\nonumber  \\
\overline B & = & f' \overline V W \gamma(R \bar v \sigma -2),
\eea
where we have introduced the ratio of derivatives
\be
R\equiv {f''(\bar\gamma \gamma) \over f'(\bar\gamma \gamma) }.
\ee
Positivity of the second variation is determined by the positivity of
\be
Q \equiv {A^2 - 4 \overline B B \over  4 |V|^2 |W|^2 f'^2 }.
\ee
which is given by
\bea
Q & = & |v|^2 (R |\gamma|^2 +1)^2 - |\gamma|^2(R\bar v \sigma - 2)(R v \bar\sigma -2)
\nonumber  \\
& = & 2 R |\gamma|^2 \ ( \ |v|^2
   +  v \bar \sigma
  + \overline v \sigma  \ )
   +   \ |v|^2 - 4 |\gamma|^2,
\eea
where the term in $R^2$ has cancelled because $|\sigma|^2 = |\gamma|^2$.

   To evaluate this expression, we use $v = 1 - \sigma$, which gives
 \be
 |v|^2 = (1 - \sigma)(1 - \bar\sigma) = 1 - \sigma - \bar\sigma + |\gamma|^2
 \ee  
 \be
 v \bar \sigma
  + \overline v \sigma = (1 - \sigma)\bar\sigma + (1 - \bar\sigma) \sigma = \sigma + \bar\sigma - 2 |\gamma|^2,
 \ee
so
 \be
 |v|^2 + v \bar\sigma + \bar v \sigma = 1 - |\gamma|^2,
 \ee
and we obtain
 \be
 Q = 2 R |\gamma|^2 ( 1 - |\gamma|^2) + 1 -2 |\gamma| \cos \phi + |\gamma|^2 - 4 |\gamma|^2,
 \ee
 where we have used $\sigma + \bar\sigma = 2 |\gamma| \cos\phi$.
 This expression is a minimum at $\cos\phi = 1$, so $Q$ will be positive for all $m$ and $n$ if and only if $Q$ is positive at this minimum, namely if
 \be
 \label{Qmin}
 Q_{\rm min} \equiv 2 R |\gamma|^2 ( 1 - |\gamma|^2) + (1 - |\gamma|)^2 - 4 |\gamma|^2
 \ee
is positive.  For $|\gamma|$ close to 1, all terms are small except the last one --- which is negative --- unless we can save the day by an appropriate choice of $f(|\gamma|^2)$.  Indeed let us choose
 \be
 \label{fofx}
 f(x) = { 1 \over (1 - x)^p },
 \ee  
where  $p$ is a power at our disposal.  We have 
\be
R(x) = {f''(x) \over f'(x) } =  { p + 1 \over 1 - x }
\ee
and, with $x = |\gamma|^2$, we obtain
\bea
 Q_{\rm min} = 2 (p + 1) |\gamma|^2 + (1 - |\gamma|)^2 - 4 |\gamma|^2
 \nonumber \\
 = 2 (p - 1) |\gamma|^2 + (1 - |\gamma|)^2.
 \eea
This will be positive for all $|\gamma| \leq 1$ if and only if $p \geq 1$.  Thus for $f(x)$ of the form \eqref{fofx}, $Q$ is non-negative for all $|\gamma| \leq 1$ and all integers $m$ and $n$ provided that 
\be
p \geq 1.
\ee
This is necessary and sufficient for $\delta^2 F$ to be a positive form.  Other expressions for $f(x)$ will also satisfy this condition, but they must have the singularity at $|\gamma| = 1$ of the strength found here.  For example $f(x) = -  \ln(1 - x)$ will not do, and  
with the simplest choice $F=\int dz d\z \ \mu\bar \mu$, one would gets a negative eigenvalue for $|\gamma| >1/3$. 

  The condition 
\be
Q_{\rm min} \geq 0,
\ee
where $Q_{\rm min}$ is defined in \eqref{Qmin} provides a simple criterion which determines whether the second variation of the minimizing functional $\delta^2 F$ is a positive form or not.

%
%
%
%
%
%
%
 \section{Definiteness and convergence of the gauge-fixing process through stochastic quantization}
 
 Stochastic quantization materializes quantum fluctuation by a Langevin equation, with a Gaussian noise $b$ and a drift force that is equal to the sum of the classical equation of motion, $- { \delta S\over \delta \phi}$ and a ``force", $\delta_v(\phi)$, tangent to the gauge orbit, that is given by a gauge transformation (in our case a reparametrization) with a field-dependent generator $v^z$. The latter must be chosen in such a way that the Langevin process converges at infinite values of a  stochastic time $t$, and the Langevin equation for any given field reads, in general 
 \bea
 \partial_t  \phi   = - \frac{\delta S}{ \delta \phi}^{\rm classical} + \delta_{{\rm gauge}, v}  (\phi) + b_\phi(x, t),
 \eea
where $b_\phi (x,t) $ is a white noise for $\phi (x,t)$.   The correlation functions of gauge independent operators cannot  depend on the the choice of $v$ (provided the stochastic process is well defined).

In the case of 2d-gravity, the last equation remains formal because in order to achieve the condition $|\mb|<1$, one cannot assume  stricto-sensu that all fluctuations of the noise $b_\phi(x, t)$ are allowed. This problem is possibly solved by reformulating the stochastic process under the form of Fockker--Planck equation, where the notion of a noise disappears when the  Fockker--Planck kernel is introduced.

       For our case, the gauge symmetry is 2d-reparametrization.  All fields now depend  on $z,\bz,t$, and  for every gauge orbit,  we introduce the following metric dependent gauge function 
 \bea
 v^z  =\rho^{z\z}f' \nabla ^{(\mbb)}_ z \mb
 \eea
 Call   $b_{\mb}$ and $b_X$  Gaussian noises for 
 $ {\mb}$ and $ X$. Both Langevin equations for the Beltrami differential and the string field are 
 \bea
 \partial_t   \mb  = T_{zz}  -  \nabla _\bz (\rho^{z\z}f' \nabla ^{(\mbb)}_ z \mb)  +b_{\mb}
 \eea
where $T_{zz} $ is the classical energy momentum tensor 
 \bea 
 T_{zz}  = \frac{\delta S}{ \delta \mb}^{\rm Polyakov }= \frac{\p_\z   - \mb \p_z   }{1-\mb\mbb}  X\cdot  \frac{\p_\z  - \mb \p_z  }{1-\mb\mbb}  X
 \eea
 and 
  \bea
 \partial_t   X  = -(\nabla _\bz \nabla _z +  \nabla _z \nabla _\bz )  X  +  \rho^{z\z}f' \nabla ^{(\mbb)}_ z \mb \nabla_ z  X  + \rho^{z\z}f' \nabla ^{(\mb)}_ \z \mbb \nabla_\bz  X+b_X
 \eea
 The presence of a   Laplacian  with no zero modes in both equations ensures  a well-defined converging stochastic process, and the gauge-fixing is well-achieved  in this method. To implement the form of the explicit equilibrium Fokker--Planck distribution of the Langevin process is  probably an impossible task, since both Langevin equations involve  nontrivial gravitational interactions between the $\mb$ and $X$ fields 
 having explicitly no zero-mode problems in the stochastic process, but a ghost-free field theory has a price, namely   the existence of of gravitational interactions.  
 
 The role of the functions $\ln   \rho_{z\z}(z,\z) $ and $f$ in the definition of the drift force along the gauge orbit  is to ensure that the latter is always a restoring one, and that it can  vanish only at the boundary  of a fundamental domain.
  If these functions are   not well chosen,   an artificial singularity of the Langevin/Fokker--Planck process may occur, where the drift force can change sign, but this just an artifact of a bad system of coordinates, which is analogous to the (pseudo) Schwartzschild singularity in the description of a black hole.

\section{Conclusion}
This paper high-lights the property that the   Gribov question is not a problem in string theory. There is  an   unambiguous  gauge-fixing, with a minimizing principle on each orbit, such that 
  the Faddeev-Popov determinant in a BRST-invariant description cannot  possibly change sign in a fundamental domain.
  Infrared problems may occur for certain  modular invariant observables, when the moduli approaches the singular points of the  fundamental domain.   Their existence is certain, since  a multitorus of genus $g$ can be pinched in a   number of ways, and can be identified as  a  Riemann surface of lower genus with identified points, a geometrical feature that seems to be   the origin  of possible   IR divergencies of    the field  theory limit of string theory.

The method indicates that   a complete knowledge of the moduli space of Riemann surfaces  is  necessary to get  a reliable    BRST-invariant action for the theory.   Since the  method has a straightforward generalization for the superstring,  we left  aside    the tachyon problem, which is irrelevant for the question of gauge-fixing.

The string is thus a very interesting laboratory for   gauge-fixing questions.  Choosing an absolute minimum for a gauge-fixing functional on each orbit  selects a unique representative of the worldsheet metric, orbit per orbit. This choice can be enforced in a BRST-invariant way. It allows one to select  and compute  all observables of the theory,   while respecting all BRST Ward identities. The expressions  found are given by the usual integrals over a fundamental  domain of  Riemann surfaces, at a fixed genus.

 This fundamental domain  is in fact  found by minimizing a certain distance in the space of Beltrami differentials, which corresponds to  the gauge-fixing functional on each orbit.   

   In the case of the torus one can explicitly verify that no Gribov issue arises. A   horizon exists however, and is found to be  the boundary of the Poincar\'e disk, where the quantum-field-theory limit of string theory is defined. This boundary of the Teichm\"uller space  is degenerate, in the sense that it represents a surface for which  the  absolute minimum of the gauge-fixing functional  is degenerate. It gathers all   the singular points of  the boundaries of each fundamental domain, when the torus becomes degenerate, as a sphere with a pair of points identified (pinched torus). However, when one restricts to one given fundamental domain, only one of these points occurs, and   its contribution can  safely  regularised, provided   one  computes infra-red safe observables.

 {\bf Acknowledgements :}
 We thank L. Alvarez-Gaume, S. Cappell,  M. Douglas,  S.  Grushevsky, E. Y. Miller, N. Nekrasov,   I.M. Singer and
L. Takhtajan for very interesting discussions.  We are grateful to M. Porrati  and  R. Stora for many valuable and informative conversations.

\section{  Appendix  A : Sketch of the   condition  $\nabla_{\bar z} h_{zz}$ in a standard BRST construction}


In this section,   for the sake of curiosity, we   show an attempt to directly enforce the minimizing gauge-condition $\nabla_{\bar z} h_{zz}$ in  the  ``standard" BRST construction, as one does in the  perturbative Yang--Mills landau gauge. For this purpose,  one  uses    a Lagrange multiplier local field $\lambda_z$ for imposing the condition by adding to the action  a term
$\int d^2z \lambda_z \nabla_z\mb$. To make this term part of a BRST-exact term, one also introduces an    anti-ghost field  $C_z $, such that 
$sC_z = \Lambda_z$.   One does the analogous for the other  sector.

The   anti-ghost $C_z $ cannot have generic zero modes, since it has a single holomorphic index, like  the Faddeev Popov ghost $c^z$, but the existence  of the 3g-3 global zero modes will pop up in a different manner as for the antighost $b_{zz}$ of the previous method. These zero modes will be carried by the now propagating Beltrami differential, and a deficit between the number of propagating  zero modes of the Beltrami differential and the Lagrange multipliers.   The theory seems  in fact almost impossible to solve, since we  will get a theory where the 2d-gravity fields  become propagating, apparently like   the longitudinal gluon  in the Yang--Mills theory in the Landau gauge.  


According to the ``naive" idea of BRST quantization, we thus 
  tentatively def{i}ne the BRST gauge-fixing action   action as 
\bea
\int d^2z\ 
   s   \Big ( C^z     (\dzb -     \mb   \dz -   2  \dz\mb        ) f'      \rho_{z  \z}     \mbb 
+s  (   C^\z   (\dz -     \mbb   \dzb -   2  \dzb\mbb        ) f'      \rho_{z  \z} \mb \Big )\CR
 = \int d^2z\ 
s   ( \Cz  \nabla _\bz  f'      \rho_{z  \z}    \mbb  
+    \Czb    \nabla _z   f'      \rho_{z  \z}   \mb ) \ \ \ \ \ \ \ \   \ \ \ \ \ \ \ 
\eea
that is, 
\bea
\int d^2z\ 
   ( \Lambda^z   \nabla _\bz    f' \rho_{z  \z}     \mbb
+    \Lambda^\z    \nabla _z      f' \rho_{z  \z}  \mb 
-  \begin {pmatrix}C^z  & C^\z \end {pmatrix}
\begin {pmatrix}
 s (\dzb -     \mb   \dz -   2  \dz\mb        ) f'  \rho_{z\z} \mbb \\  s (\dz -     \mbb   \dzb -   2  \dzb\mbb        )     f' \rho_{z  \z}  \mb   
\end {pmatrix}
\CR
\eea
This action is problematic. The ghost terms are probably well defined by a proper choice of the function $f'$. However, one has  global zero modes for  $f' \rho_{z  \z}\mb$ and $f' \rho_{z  \z} \mbb$.
One must  force   $f'\rho \mu$    to remain in the appropriate space   of the same dimension as 
 $\Lambda$, by a gauge-fixing involving constant ghosts. This is  probably the way  an   integration over a fundamental  domain  will make its way in the expression of the partition function.   There is not much motivation to check the details because in this action the Beltrami differentials now become propagating fields, as do
 $ \Lambda_\bz  $ and  $\Lambda_z$, and one gets gravitational interactions  with    the string field $X$. We thus expect   super-renormalizable   2-d quantum field theory, with  a subtle infra-red problem.\footnote{ This QFT  has a chance to be handled in the limit of infinite genus, which is unreachable in the normal construction, because of the growing  complicated structure of fundamental domains when the genus increases.}

 One can  however  check  the   consistency of this theory by computing its conformal anomaly, which only involves    the local structure of the worldsheet. This is a purely local question that   can be done at genus zero.  One must     compute perturbatively   $\nabla_\bz T_zz(x) ,  T_zz(y) $ and check its vanishing condition, to be able to enforce the BRST Ward identity. This  computation was   done a long time ago, (it was motivated by different concerns \cite{bilal}). The computation with a propagating metric  involves loops containing the free propagators of $\mu, \Lambda, c, C$ and $X$. The contribution of the ghosts is not the same as in the conformal gauge, due to the different conformal weights of the anti-ghosts, but one still gets the condition $D-26=0$ due to compensating contribution of internal loops   of  $\mb  $ and~$\Lambda$.
 
   It is important that the method we advocate of defining the gauge-fixing by the minimizing principle on each orbit is however completely well defined, since, as shown in this Appendix,  the attempt to   enforce the condition $\nabla f'\rho \mu=0$  in a conventional BRST-invariant way leads to unnecessary   stringy complications, such as the  propagation of lagrange multipliers of the BRST symmetry, with the occurrence of extra zero modes that seem difficult to solve.

\section{  Appendix  B:  Superstring extension}

For the superstring case, the Beltrami differential  gets a supersymmetric partner, the conformal invariant  part of the 2d gravitino, with  2 components   $\alpha_z^+(z,\z),\alpha_\z^-(z,\z)$. 
The 2d spinor $\alpha_\z^+,\alpha_z^-$ is  defined in the tangent plane of the Riemann surface, and its large gauge transformations are deduced from those of the Beltrami differential.  Calling $\gamma^\pm$ the local ghost of local supersymmetry, the  small reparametrization and supersymmetry gauge transformations are represented by the following BRST transformations
  \bea \label{brsmualpha}
  s \mb &=&\dzb \cz   +\cz \dz \mb - \mb \dz  \cz   + \alpha_\z^+ \gamma^z \CR s \cz&=&  \cz\dz\cz +\frac{1}{2}\gamma^z \gamma^z 
  \CR
   s \alpha_\z^+&=&\dzb \gamma^z +\mb \dz  \gamma^z - \frac{1}{2}\gamma^z \dz  \gamma^z  
    +\cz \dz \alpha_\z^+- \frac{1}{2}\alpha_\z^+ \dz  \cz    
    \CR
    s \gamma_z^+&=&\cz \dz \gamma_z ^+- \frac{1}{2} \gamma_z ^+ \dz  \cz 
  \eea
and complex conjugate equations. The question of the gauge-fixing of the local supersymmetry can be solved with the  generalisation of the   method we introduced for the Beltrami differential. There are gauge orbits for $\alpha_\z^+$ and $\alpha_z^-$. The choice of a unique representative both for $\mu$ and $\alpha$ will be obtained by a minimizing principle, using for instance the functional
\bea \label{extreme} 
F [\mb, \mbb, \alpha_z, \alpha_\z] =\int_\Sigma  \rho_{z\z} (z,\z)   dzd\bz   \  \frac {1} {{1-|\mu(z,\z)| ^2}} \\   
\Big (   \ln \frac {1-|\mu(z,\z)|}{1+|\mu(z,\z)|}  -\sqrt {\rho^{z\z}} \alpha_\z^+(z,\z)\alpha_z^-(z,\z)
\Big).
\eea
For instance, at genus one, the solution  for the minimum  is  $\mu^z_\z =\gamma$ and  $\alpha^+_\z =t$, where $t$ is a super-module, and for   genus $g > 2$, the Riemann-Roch theorem predicts the integration over 2g-2 super-modules, with a method completely analogous as the one we followed for the Beltrami differential, and an eventual partition with a BRST symmetry. In the path integral, the super-module is a Grasmann variable, and its BRST transform is a commuting  constant $T$,  with $st=T$. $T$ is unbounded  and serves as a bosonic constant Lagrange multiplier for ensuring that the commuting antighost $\beta_ \z^- $ has no zero modes.

\end{document}